\documentclass[showpacs,aps,pra,twocolumn,amsmath,amssymb]{revtex4}

\usepackage{mathbbol}
\usepackage{graphicx}
\usepackage{color} 
\usepackage{cancel}
\usepackage{ulem}
\usepackage{soul}
\definecolor{orange}{rgb}{1,0.5,0}

\newcommand{\ket}[1]{|#1\rangle}
\newcommand{\bra}[1]{\langle #1|}

\begin{document}

\title{Maximally discordant mixed states of two qubits}

\author{Fernando Galve}
\author{Gian Luca Giorgi}
\author{Roberta Zambrini}
\affiliation{IFISC (UIB-CSIC),
Instituto de F\'isica Interdisciplinar y Sistemas Complejos, UIB Campus,
E-07122 Palma de Mallorca, Spain}

\date{\today}
\begin{abstract}
We study the relative strength of classical and quantum correlations, as measured by discord,
for two-qubit states.  Quantum correlations appear only in the presence of classical correlations, while
the reverse is not always true.  We identify the family of  states that maximize the discord for
a given value of the classical correlations and show that the largest attainable discord for mixed
states is greater than for  pure states.  The difference between discord and entanglement is
emphasized by the remarkable fact that  these states do not maximize entanglement and are, in
some cases, even separable.  Finally, by random generation of density matrices uniformly
distributed over the whole Hilbert space, we quantify the frequency of the appearance of quantum
and classical correlations for different ranks.

\end{abstract}

\pacs{03.67.Mn, 
03.65.Ta} 

\maketitle

\section{Introduction}
One of the most striking features of quantum mechanics is entanglement, first considered (although not by
that name) by Einstein, Podolsky, and Rosen in their seminal paper in 1935 \cite{epr}. This is an exclusively
\textit{quantum} feature of composite states that can not be written as mixtures of product states. Theoretical and
experimental research activity to characterize entanglement has been particularly intense in the last
decade (see review \cite{horo2009} and references therein), 
being part of a broader endeavor to explore distinctive aspects of
quantum versus classical physics and novel resources for quantum information purposes \cite{nielsen}. An
important issue considered by several authors  \cite{zurek,vedral,horods}
 is the existence of quantum correlations \textit{beyond} entanglement, in separable states. 
As a matter of fact, examples of
improved quantum computing tasks \textit{not} relying on entanglement have been reported
\cite{info-no-ent}.

\section{Quantum correlations: the discord} 
Two complementary approaches on quantum correlations are receiving great attention \cite{zurek,vedral}. 
In Ref. \cite{zurek}, quantum correlations (quantum discord) have been associated to the difference of 
two classically equivalent  expressions for the mutual information, $\cal{I}$ and $\cal{J}$. In
particular, the quantum mutual information is defined as
${\cal{I}}(\varrho)=S(\varrho_A)+S(\varrho_B)-S(\varrho)$, where $S$ stands for the von Neumann entropy
and $\varrho_{A(B)}$ is the reduced density matrix of each subsystem. The classically equivalent
expression stemming from Bayes rule is ${\cal{J}}(\varrho)_{\{\Pi_j^B\}}=S(\varrho_A)-S(A|\{\Pi_j^B\})$,
with the conditional entropy defined as $S(A|\{\Pi_j^B\})=\sum_ip_iS(\varrho_{A|\Pi_i^B})$, $p_i={\rm
Tr}_{AB}(\Pi_i^B\varrho)$, and where $\varrho_{A|\Pi_i^B}= \Pi_i^B\varrho\Pi_i^B/{p_i} $ is the density
matrix after a complete projective measurement $(\{\Pi_j^B\})$ has been performed on B. 
Quantum discord is obtained minimizing the difference    ${\cal{I}}(\varrho)-{\cal{J}}(\varrho)$:
\begin{equation}\label{eqdisc}
\delta_{A:B}(\varrho)=\min_{\{\Pi_i^B\}}\left[S(\varrho_B)-S(\varrho)+S(A|\{\Pi_i^B\})\right],
\end{equation}
that is, when measurement is performed in the basis which disturbs the state the
least. 
A complementary approach was described in Ref. \cite{vedral}, defining  classical correlations and  showing
that total correlations given by the mutual information are actually  larger thAn the sum of the
classical correlations and entanglement $E$ \cite{wootters}. As a matter of fact, the quantum mutual information can be seen as
the sum of quantum correlations \cite{zurek} $\delta_{A:B}(\varrho)$ and the classical correlations \cite{vedral}
$\max_{\{\Pi_i^B\}}{\cal{J}}(\varrho)_{\{\Pi_j^B\}}$. We remark that, in \cite{zurek}, the discord is defined
in terms of orthogonal (perfect) measurements. Even if possible generalizations to positive-operator-valued
measurements (POVM) were considered at the end of that paper, as well as in \cite{vedral}, calculations of discord in the
literature generally consider only orthogonal measurements (see, e.g., \cite{zurek,info-no-ent,maziero,luo2008,mazhar2010,mazzola2010,guo}).

In contrast with state separability, this new paradigm of quantumness of correlations is 
measurement oriented, considering an experiment where all information of a system A is extracted by
measuring another system B. According to this measure, a state is
classically correlated only when consecutive measurements of system B yield the same picture of the
state of system A, which is achieved after decoherence into the pointer basis of B \cite{zurek,maziero}.
For pure states, quantum discord is equivalent to entanglement and
 actually has the same value of classical correlations \cite{vedral}. 
On the other hand, when mixed states are considered,  entanglement does
 significantly depart from the quantum discord, the difference being positive for some states and negative for others \cite{luo2008}.
As the definition of discord comes
from a minimization over all possible measurement basis, only a few general results  have been
reported.   Analytic expressions are known for states of two qubits with maximally mixed marginals
\cite{luo2008}, for X-shaped states \cite{mazhar2010}, and also for Gaussian states of continuous
variable systems \cite{paris-adesso}. 

In this paper, we explore the whole Hilbert space of two qubits to gain insight on their
correlations for mixed states of different ranks. Our main goal is to discern the 
proportion of quantum to classical correlations between the two qubits.
We find the most nonclassical two-qubit states, i.e., the family with maximal quantum
discord versus classical correlations, were formed by mixed states of rank 2 and 3, which we name
maximally discordant mixed states (MDMS). The analogous effort to identify largest
deviations from classical states has led to 
 mixed states maximizing
entanglement versus purity  \cite{MEMS}. In contrast with maximally entangled mixed states (MEMS), where a geometrical
property such as separability could be considered with several constraints, 
the MDMS are naturally defined and allow us to quantify the relative strength
of quantum and classical correlations, which are related in a closed from. 
 The MDMS proposed here do not maximize entanglement 
for a given amount of classical correlations; part of them are, in fact, separable. This pinpoints
 the fundamental difference between entanglement and discord for mixed states, in opposition to their
exact equivalence for pure states. 
Their discord also implies that quantum correlations are always accompanied by classical correlations,
while the reverse is not always true. Furthermore, we study the probability of states with a given amount of discord in the whole two-qubit
Hilbert space, supporting the recent result that the closed set of purely classically correlated
states ($\delta_{A:B}=0$) has measure zero \cite{acin2010}, and we compare it with the probability for
classical correlations and entanglement. Only such states with no discord have been shown to ensure a
future non-negative evolution in the presence of dissipation \cite{shabani2009}, while discord can not be made
zero in a finite time by any Markovian map \cite{acin2010}.  
Furthermore, even in the presence of a noisy environment, for some family of initial states, 
discord can be robust under
decoherence for a finite time \cite{mazzola2010}. Experimental results with polarization
entangled photons also have been reported recently \cite{guo}.

\section{Mixed states with largest discord} As pure states with maximum entanglement that are also maximally
discordant, Bell states are a natural starting point to identify states with large discord; thus, by
mixing these states with other components, it is quite plausible that we find states with a large
proportion of quantum versus classical correlations. 
We then consider an example of a mixture of any Bell state  $\ket{\psi}$ with another orthogonal pure
state, i.e., $\varrho=\epsilon\ket{\psi}\bra{\psi}+(1-\epsilon)\ket{\phi}\bra{\phi}$. If $\ket{\phi}$ is
any other Bell state, then $\delta_{A:B}=E$  (with $E$ being entanglement) and $\mathcal{J}=1$, and  a
worse discord is found than in that of pure states with the same classical correlation $\mathcal{J}$.  In
contrast, a mixture of a Bell state and a state of the computational basis of the opposite parity
sector, gives a huge amount of discord. We thus consider states
\begin{equation}\label{symmRank2}
\varrho=\epsilon\ket{\Phi^+}\bra{\Phi^+}+(1-\epsilon)\ket{01}\bra{01} 
\end{equation} 
with $\ket{\Phi^+}=(\ket{00}+\ket{11})/\sqrt{2}$ as usual. With local unitary operations, which leave
discord invariant, we can obtain from this ansatz any combination of a Bell state mixed with a
computational basis state of opposite parity (number of 1's in the state). The expression of discord for
 states (\ref{symmRank2}) is invariant under permutation of the individual labels $A\leftrightarrow B$. As a matter
of fact, we find that states (\ref{symmRank2}) maximize the symmetrized version of  discord
$(\delta_{A:B}+\delta_{B:A})/2$  for all rank-2 matrices.

\begin{figure}[h]
\includegraphics[width=8cm]{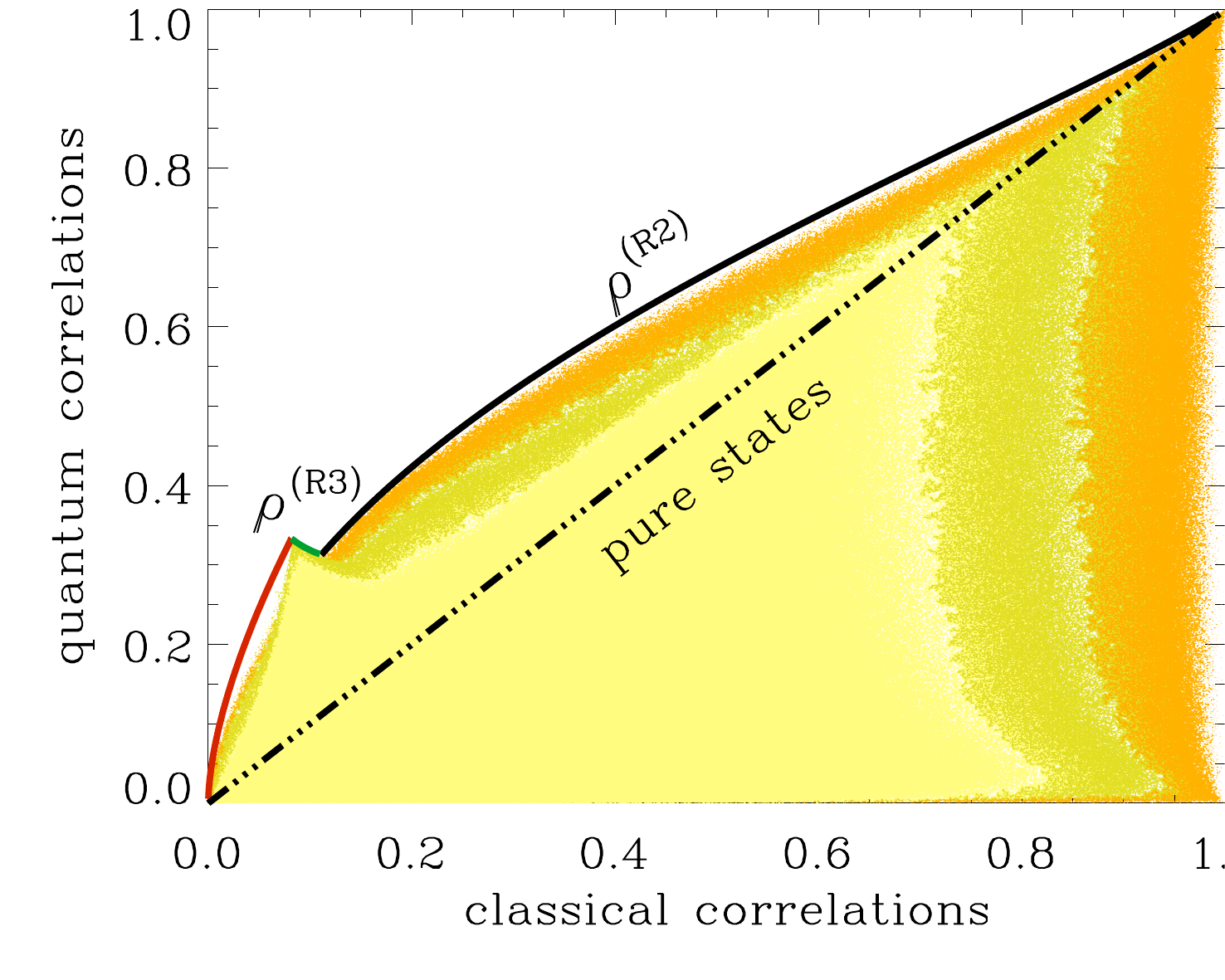} \caption{(Color online) Quantum discord ($\delta_{A:B}$) versus
classical correlations  ($\mathcal{J}$) for two-qubit states.  The MDMS family (continuous line) gives
two segments for rank 3 ($\varrho^{\text{(R3)}}$) and one for rank 2 ($\varrho^{\text{(R2)}}$). Layers
of $10^8$ random matrices of rank 2 (dark points), 3 (intermediate color), and 4  (lighter color) are
superimposed.  For pure states (dotted-dashed line), $\delta_{A:B}=\mathcal{J}=E$.\label{fig1}}
\end{figure}

It can be shown that, when we compare with a numerical scan of Hilbert's space, this state is too symmetric. 
In fact, a better option in terms of discord is obtained if some amount of entanglement is sacrificed for the good of 
quantum correlations. This results from asymmetrizing the maximally entangled (Bell) state leading to the ansatz
\begin{equation}\label{asymmRank2}
\varrho^{\text{(R2)}}=\epsilon\ket{\tilde{\Phi}^+}\bra{\tilde{\Phi}^+}+(1-\epsilon)\ket{01}\bra{01},
\end{equation}
with $\ket{\tilde{\Phi}^+}=\sqrt{p}\ket{00}+\sqrt{1-p}\ket{11})$, which coincides with a Bell state for $p=1/2$.
The increase of discord for the states (\ref{asymmRank2}) with respect to (\ref{symmRank2}) highlights the importance of the asymmetric definition of quantum discord, 
based on the asymmetric operation of measuring B in order to know about A.
The discord for this family can be written once we know the conditional entropies min$_{\{\Pi_i^B\}}S(A|\{\Pi_i^B\})=(x\log_2\frac{1-x}{1+x}-\log_2y)/2$, with $x=\sqrt{1-4y}$ and $y=\epsilon(1-p)(1-\epsilon)$, while for $\delta_{B:A}$ we need to use $y=\epsilon p(1-\epsilon)$. The total and reduced entropies are easy to calculate once we notice that the ansatz is given in spectral decomposition, though we do not give the whole expression for reasons of space.

The family of MDMS is obtained for an optimal function $\epsilon_{\text{opt.}}(p)$ through the use of Lagrange multipliers,
 as detailed later in this paper. Once this optimal curve $\epsilon_{\text{opt.}}(p)$ is used, Eq. 
(\ref{asymmRank2}) gives the family of states that maximize the quantum part of correlations for a given classical part
, when all rank 2 (R2) states are considered. As shown in Fig. \ref{fig1}, these $\varrho^{(\text{R2})}$ states are the MDMS for a large range of classical correlations.

\begin{figure}[h]
\includegraphics[width=8cm]{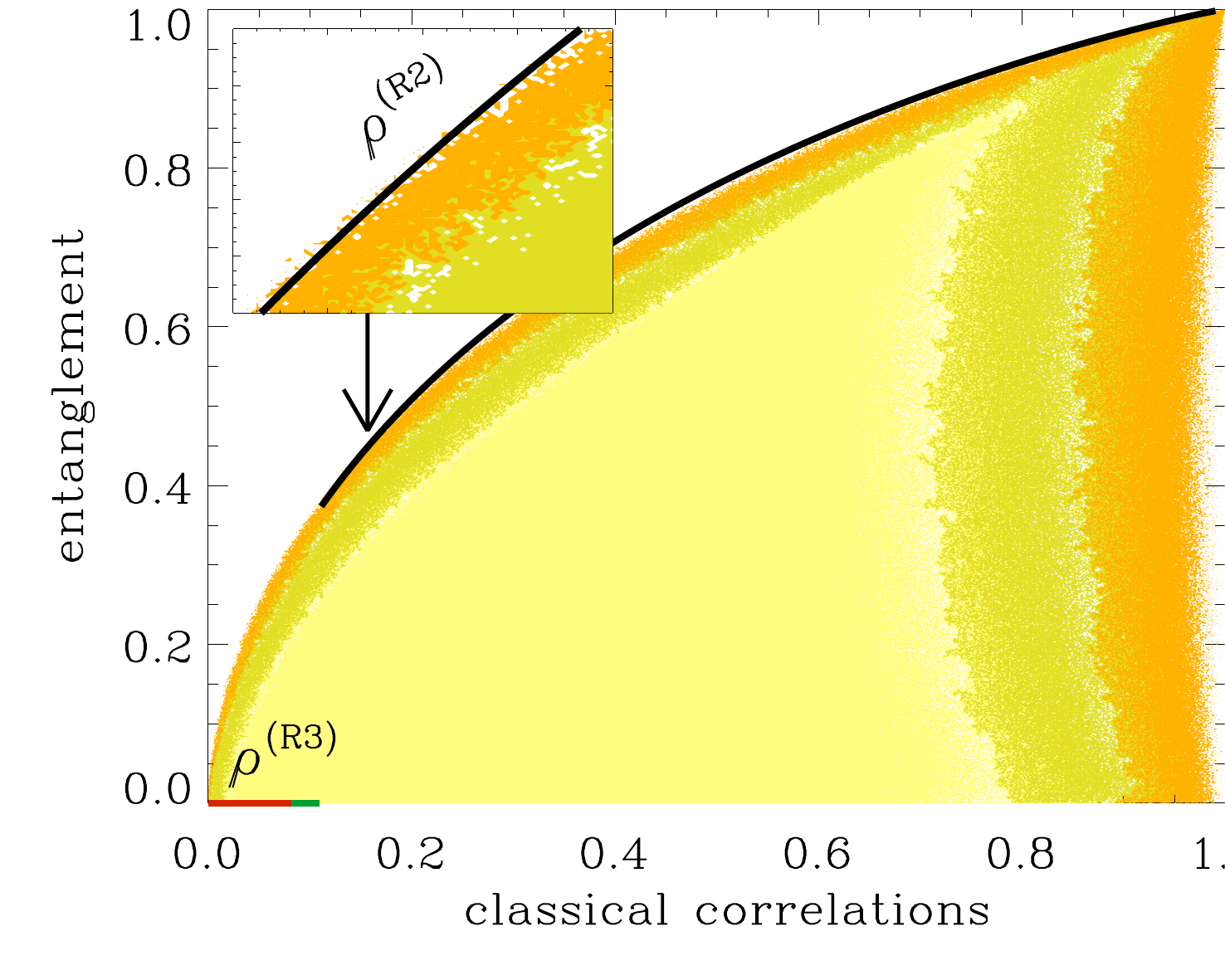}
\caption{(Color online) Entanglement ($E$) vs classical correlations  ($\mathcal{J}$). Random matrices 
and lines for the MDMS family as in Fig. \ref{fig1}.
MDMS of rank 3 are separable, while $\varrho^{\text{(R2)}}$ states show large entanglement, even if not
maximum, as seen
in the inset. \label{fig2}}
\end{figure}

In order to find the states that maximize $\delta_{A:B}$, we also need to use rank-3 states. 
In this case, it can be checked that asymmetrization of the Bell-state component $\ket{\phi}$ does not help, and that the best choice 
for an ansatz is
\begin{equation}\label{symmRank3}
\varrho^{\text{(R3)}}=\epsilon\ket{\Phi^+}\bra{\Phi^+}+(1-\epsilon)(m\ket{01}\bra{01}+(1-m)\ket{10}\bra{10}).
\end{equation}
As before, any combination of Bell state plus two components of opposite parity belonging to the
computational basis, will do. The optimal $\epsilon_{\text{opt.}}(m)$ is discussed later and leads to
the family of states $\varrho^{(\text{R3})}$ maximizing discord for small classical correlations. This
optimal family has the property of being separable (not entangled), 
as shown in Fig.~\ref{fig2}, while it maximizes quantum discord,
highlighting the inequivalence of these measures of quantumness. It is
actually found that, for these states, the discord amounts to the weight of the Bell component, and the
simple relation $\delta_{A:B}=\delta_{B:A}=\epsilon$ holds. For completeness, the entanglement (as
quantified by the concurrence \cite{wootters}) yields $E(\varrho^{(\text{R2})})=2\epsilon\sqrt{p(1-p)}$
and $E(\varrho^{(\text{R3})})=\max [0,\epsilon-2(1-\epsilon)\sqrt{m(1-m)}]$.
Although MDMS of rank 3
are separable, the MDMS of rank 2  have a high amount of entanglement, 
even if they do not maximize it (Fig.~\ref{fig2}). As mentioned before,
 asymmetrization of the Bell component increases quantum correlations at the expense of
entanglement. 

 An intriguing state is that of 
the singular point for $\varrho^{(\text{R3})}$ shown in Fig.~\ref{fig1}:
$\varrho^{\text{cusp}}=(\ket{\Phi^+}\bra{\Phi^+}+\ket{01}\bra{01}+\ket{10}\bra{10})/3$ reaches the lowest
possible purity for a rank-3 state and is separable, yet with a high level of discord.
Another important feature emerging from Fig.~\ref{fig1} is that 
MDMS have a discord larger than pure states 
($\delta_{A:B}>\mathcal{J}$)  and satisfy $\mathcal{J}=0$ only when $\delta_{A:B}=0$, 
thus showing the lack of states with finite quantum without classical
correlations. In other words, no state of two qubits is purely quantum. 
Maybe less surprisingly, there are no states with finite entanglement and zero classical
correlations (Fig.~\ref{fig2}). 

\section{Detailed analysis}
 We first consider the commonly accepted definition of discord, obtained expressing the measurement projectors as $\Pi_j^B=\ket{\psi_j^B}\bra{\psi_j^B}$, $j=1,2$, with 
$\ket{\psi_1^B}=\cos\theta\ket{0}+e^{i\phi}\sin\theta\ket{1}$ and
$\ket{\psi_2^B}=-e^{-i\phi}\sin\theta\ket{0}+\cos\theta\ket{1}$ \cite{zurek}.
The angles minimizing Eq. (\ref{eqdisc}) for the states (\ref{asymmRank2}), and thus giving the correct discord, are
$\theta=\pi/4+n\pi/2$ and any value of $\phi$. We then can choose $\phi=0$ and hence the optimal
projectors are $\Pi_1^B=\ket{+^B}\bra{+^B}$ and $\Pi_2^B=\ket{-^B}\bra{-^B}$, with
$\ket{\pm}=(\ket{0}\pm\ket{1})/\sqrt{2}$. 

The method of Lagrange multipliers allows us to maximize first the discord for the rank-2 family
$\delta_{A:B}(\varrho^{\text{(R2)}})\equiv\delta(\epsilon,p)$ while keeping its classical correlations
${\mathcal{J}}(\varrho^{\text{(R2)}})\equiv \mathcal{J}_0 (\epsilon , p)$ constant (and the same for the rank-3 family
$\varrho^{\text{(R3)}}$). This is achieved through definition of the function
$\Lambda(\epsilon,p,\lambda)=\delta(\epsilon,p)+\lambda(\mathcal{J}(\epsilon,p)-\mathcal{J}_0)$ where
$\lambda$ is the Lagrange multiplier, and $\mathcal{J}_0$ is an arbitrary but fixed amount of classical
correlation. The extremization procedure is then simply the simultaneous solution of the three equations
$\partial_{\mu}\Lambda=0$, with $\mu=\lambda,\epsilon,p$. From the first equation,
$\mathcal{J}(\epsilon,p)=\mathcal{J}_0$ is obtained, as expected. 
From the last two equations, we can isolate $\lambda$ yielding the extremality condition
\begin{equation}\label{eq5}
\partial_\epsilon\delta/\partial_\epsilon \mathcal{J}=\partial_p\delta/\partial_p \mathcal{J}.
\end{equation}
We stress that this condition is equivalent to maximization of $\delta$ versus $\mathcal{I}$,  or
minimization of $\mathcal{J}$ with respect to $\mathcal{I}$, due to the closed relation
$\mathcal{I}=\delta+\mathcal{J}$. These quantities present nontrivial trigonometric relations, leading
to a transcendental equation the solution of which can only be given numerically. 

The same procedure is followed for the rank-3 family $\varrho^{\text{(R3)}}$, with $m$ playing the role of $p$. 
In this case, obtaining $\epsilon_{\text{opt.}}(m)$ is a bit
trickier,  due to the fact that there are two optimal angles (each of them good for different ranges of $\epsilon$ and
$m$), $\theta=0,\pi/4$; the angle $\phi$ is again non important. We can consider for the moment that projector maximization
of discord has been simplified to $\delta_{A:B}(\varrho^{\text{(R3)}})=\min(\delta_0,\delta_{\pi/4})$, the latter being
functions of $\epsilon$ and $m$.  The goal is to find the zero(es) of the function
$\partial_\epsilon\delta/\partial_\epsilon \mathcal{J}-\partial_m\delta/\partial_m \mathcal{J}$, which of course needs the
knowledge of when to use one angle or the other. However, the problem is greatly reduced by noticing that the latter function is
positive when using $\delta_0$ and negative when using $\delta_{\pi/4}$. This means that the zero of such function occurs
exactly (and conveniently) when $\delta_0(\epsilon,m)=\delta_{\pi/4}(\epsilon,m)$. 
 Again, the solution to this transcendental equation can only be given numerically.

Finally, the MDMS are a family of states
\begin{equation}\label{MDMS}
\varrho^{\text{MDMS}}=\epsilon\ket{\tilde{\Phi}^+}\bra{\tilde{\Phi}^+}+(1-\epsilon)(m\ket{01}\bra{01}+(1-m)\ket{10}\bra{10}),
\end{equation}
where the optimum choice of parameters gives the three curves in Fig.~\ref{fig1}, two of them rank 3 and the other
rank 2. The first curve, going from zero discord up to the cusp, is the
rank 3 family $\varrho^{\text{(R3)}}$ with $\epsilon_{\text{opt.}}(m)$ given by the solution of
$\delta_0=\delta_{\pi/4}$. It is restricted to the domain $m\in [0,1]$, $\epsilon\in[0,1/3]$. The second
branch of MDMS is given by $\varrho^{\text{(R3)}}$ with $m=1/2$ with
domain $\epsilon\in[1/3,0.385]$, approximately. These two curves correspond to separable states, as
noted above (Fig.~\ref{fig2}). The remaining curve of MDMS is the rank 2 
family $\varrho^{\text{(R2)}}$ when the
optimal function $\epsilon_{\text{opt.}}(p)$ given by Eq. (\ref{eq5}) is used, and for $\epsilon$ approximately in
the interval $[0.408,1]$.  One might wonder how the picture changes if more general (nonorthogonal) measurements are
considered. It has been shown that, for two qubits, discord is extremized exclusively by rank-1 POVMs with a maximum of four elements \cite{PhDdatta,dariano}.
 Perfect orthogonal measurements correspond to the case with the two elements considered above. Considering POVMs with
measurement operators $E_i$, the measured density matrix takes the form $\varrho_{A|E_i}=E_i\varrho/p_i$ with probability $p_i={\rm Tr}_{AB}(E_i\varrho )$. Even using the general measurement given by POVMs of four elements, we find the same discord for the MDMS, meaning that
they represent the absolute border of maximally nonclassically correlated states of two qubits. 
A detailed analysis about the full Hilbert space will be presented elsewhere \cite{mdms2}.

\section{Statistics} 
\begin{figure}[h]
\includegraphics[width=8cm]{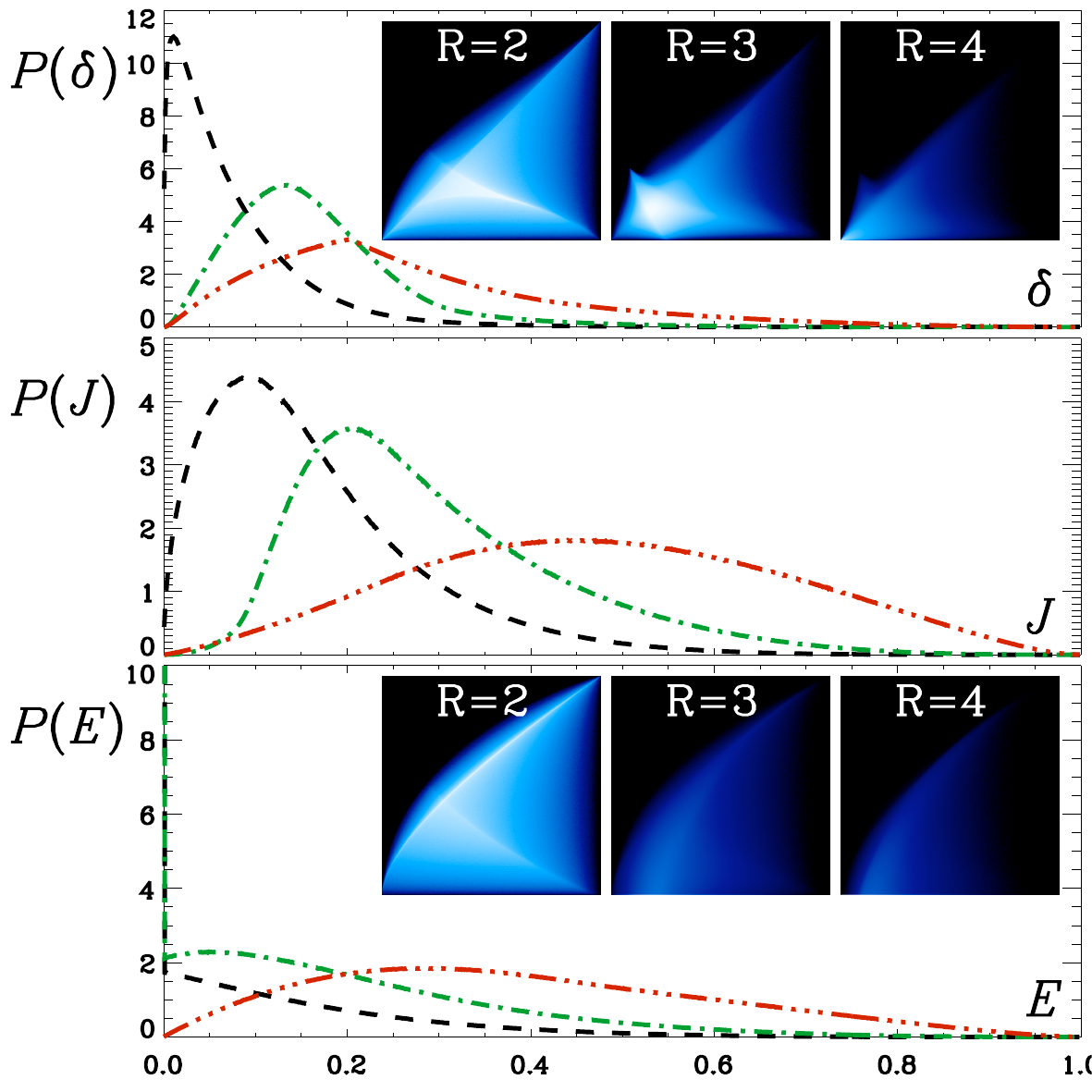}
\caption{(Color online) Probability (density) to find a two-qubit state with a given amount of quantum discord $\delta$,
classical correlations $\mathcal{J}$, and entanglement $E$ respectively, for ranks 2 (dashed),
3 (dot-dashed), and 4 (dot-dot-dashed). The insets show these probabilities (larger for light color) for the quantity under
study  ($y$ axis) against classical correlations $\mathcal{J}$ ($x$ axis), for different ranks.\label{fig3}}
\end{figure}

Since our random generation of density matrices is uniform in the Hilbert space, 
preserving the Haar measure \cite{random,volume}, we can measure the frequency of the appearance of states with different
properties, as shown in Fig.~\ref{fig3} for different ranks.  Some main features arise for all quantities investigated:
(i) Zero correlations (be they quantum, classical, or mutual information) have zero probability. Notably, entanglement is the
exception (consistently with Ref.~\cite{volume}), where only rank-2 states have a probability zero of being separable. (ii) The lower the rank, the higher the
typical amount of correlations. This is quite understandable, since higher ranks describe more mixed states. (iii)
It is more probable to find states with more
abundance of classical rather than quantum correlations.
We note that only $\sim7.45\%$ of the two-qubit states has greater discord than classical
correlations. If restricted to lower ranks, we observe that rank 2 matrices yield $\sim10.76\%$, and rank 3 yield
$\sim16.3\%$. Finally, as shown in the insets in Fig.~\ref{fig3},
the border of MDMS seems
to be rather improbable to find in the space of two qubit states, except for the middle branch in the cusp, meaning that
such extremely nonclassical states are quite rare.

\section{Conclusions} The unique family of two-qubit mixed states  with maximal proportion of quantum discord
versus classical correlations, the MDMS in Eq.~(\ref{MDMS}), has been identified. Part of them have rank 2
and are highly, although not maximally, entangled, while the other part has rank 3 and is separable, thus providing another evidence of the
inequivalence of these two measures of quantumness. We have shown that the presence of discord  is a sufficient but not
necessary condition to have nonvanishing  classical correlations. The uniform generation of states (random states
preserving Haar measure) allowed us to find the probabilities and typical values of classical and quantum correlations, as
well as entanglement. We  verified that completely (either quantum or classical) uncorrelated states are very rare, as well
as extreme nonclassical states. The identification of MDMS, together with the ability to experimentally 
generate \cite{fanchini} and characterize \cite{guo} these states, is a key tool to 
establish the fundamental difference in  performance of quantum versus 
classical information \cite{info-no-ent}.

\acknowledgments{Funding from FISICOS (FIS2007-60327), ECuSCo (200850I047), and CoQuSys (200450E566) projects and
Juan de la Cierva program are acknowledged.}


\begin{thebibliography}{10}

\bibitem{epr} A. Einstein, B. Podolsky, and N. Rosen, Phys. Rev. {\bf 47}, 777 (1935).

\bibitem{horo2009} R. Horodecki, P. Horodecki, M. Horodecki, and K. Horodecki, Rev. Mod. Phys. {\bf 81}, 865 (2009).

\bibitem{nielsen} M. A. Nielsen and I. L. Chuang, {\it Quantum Computation and Quantum Information} (Cambridge University Press, Cambridge, U.K., 2000).

\bibitem{zurek}  H.  Ollivier and W.  H.  Zurek,  Phys. Rev. Lett.  {\bf 88}, 017901 (2001).

\bibitem{vedral} L.  Henderson and V.  Vedral,  J. Phys. A  {\bf 34}, 6899 (2001).

\bibitem{horods} M. Horodecki, K. Horodecki, P. Horodecki, R. Horodecki, J. Oppenheim, A. Sen(De), and U. Sen, Phys. Rev. Lett. {\bf 90}, 100402 (2003).

\bibitem{info-no-ent} E. Knill and R. Laflamme, Phys. Rev. Lett. {\bf 81}, 5672 (1998);
S. L. Braunstein, C. M. Caves, R. Jozsa, N. Linden, S. Popescu, and R. Schack, \textit{ibid.} {\bf 83}, 1054 (1999); 
D. A. Meyer, \textit{ibid.} {\bf 85}, 2014 (2000);
E. Biham, G. Brassard, D. Kenigsberg, and T. Mor, Theor. Comput. Sci.  {\bf 320}, 15 (2004); 
A. Datta, A. Shaji, and C. M. Caves, Phys. Rev. Lett. {\bf 100}, 050502 (2008); 
B. P. Lanyon, M. Barbieri, M. P. Almeida, and A. G. White, \textit{ibid.}  {\bf 101}, 200501 (2008); 
V. Vedral, Found. Phys. {\bf 40}, 1141 (2010). 

\bibitem{wootters} W. K. Wootters, Phys. Rev. Lett. {\bf 80}, 2245 (1998).


\bibitem{maziero} J. Maziero, L. C. C\`eleri, R. M. Serra, and V. Vedral, Phys. Rev. A {\bf 80}, 044102 (2009). 


\bibitem{luo2008} S. Luo, Phys. Rev. A {\bf 77}, 042303 (2008).

\bibitem{mazhar2010} M. Ali, A. R. P. Rau, and G. Alber, Phys. Rev. A {\bf 81}, 042105 (2010).

\bibitem{paris-adesso}   P. Giorda and M. G. A. Paris, Phys. Rev. Lett. {\bf 105}, 020503 (2010);
G. Adesso and A. Datta,  \textit{ibid.} {\bf 105}, 030501 (2010).

\bibitem{MEMS} S. Ishizaka and T. Hiroshima, Phys. Rev. A {\bf 62}, 022310 (2000);
F. Verstraete, K. Audenaert, and B. De Moor, \textit{ibid.} {\bf 64}, 012316 (2001).

\bibitem{acin2010} A. Ferraro, L. Aolita, D. Cavalcanti, F. M. Cucchietti, and A. Ac\'in, Phys. Rev. A {\bf 81}, 052318 (2010).


\bibitem{shabani2009}   C. A. Rodr\'iguez-Rosario, K. Modi, A. Kuah, A. Shaji, and E. C. G. Sudarshan, J. Phys. A {\bf 41}, 205301 (2008);
 A. Shabani and D. A. Lidar, Phys. Rev. Lett. {\bf 102}, 100402 (2009).

\bibitem{mazzola2010} L. Mazzola, J. Piilo, and S. Maniscalco, Phys. Rev. Lett.  {\bf 104}, 200401 (2010).

\bibitem{guo}J.-S. Xu, X.-Y. Xu, C.-F. Li, C.-J. Zhang, X.-B. Zou, G.-C. Guo, Nature Commun. {\bf 1}, 7 (2010).


\bibitem{random} K. Zyczkowski and M. Kus, J. Phys. A {\bf 27}, 4235 (1994).
 
\bibitem{volume} K. Zyczkowski, P. Horodecki, A. Sanpera, and M. Lewenstein,, Phys. Rev. A {\bf 58}, 883 (1998).

\bibitem{fanchini} F. F. Fanchini, L. K. Castelano, and A. O. Caldeira, New J. Phys. {\bf 12},  073009 (2010).

\bibitem{PhDdatta} A. Datta, PhD thesis, arXiv:0807.4490.
\bibitem{dariano} G. M. D'Ariano, P. Perinotti and P. Lo Presti, J. Phys. A {\bf 38}, 5979 (2005).
\bibitem{mdms2} F. Galve, G. L. Giorgi and R. Zambrini, in preparation.


\end{thebibliography}
\end{document}